\documentclass[aip,jcp,reprint]{revtex4-1}
\usepackage{amsmath,amssymb,bm,graphicx,color,multirow,longtable}

\newcommand{\alert}[1]{\textcolor{black}{#1}}

\newcommand{\bx}{\mathbf{x}}
\newcommand{\EHF}{E_{\rm HF}}

\newcommand{\Ec}{E_c}
\newcommand{\mEh}{m$E_{\rm h}$}
\newcommand{\uEh}{$\mu E_{\rm h}$}
\newcommand{\ds}{\displaystyle}
\newcommand{\mc}{\multicolumn}
\newcommand{\mr}{\multirow}

\begin{document}

%======================================================
\title{Uniform electron gases: III. Low-density gases on three-dimensional spheres}
%======================================================
\author{Davids Agboola}
\author{Anneke L. Knol}
\author{Peter M. W. Gill}
\email{peter.gill@anu.edu.au}
\author{Pierre-Fran\c{c}ois Loos}
\thanks{Corresponding author}
\email{pf.loos@anu.edu.au}
\affiliation{Research School of Chemistry, Australian National University, ACT 2601, Australia}

%==========
\begin{abstract}
By combining variational Monte Carlo (VMC) and complete-basis-set limit Hartree-Fock (HF) calculations, we have obtained near-exact correlation energies for low-density same-spin electrons on a three-dimensional sphere (3-sphere), i.e.~the surface of a four-dimensional ball.  In the VMC calculations, we compare the efficacies of two types of one-electron basis functions for these strongly correlated systems, and analyze the energy convergence with respect to the quality of the Jastrow factor.  The HF calculations employ spherical Gaussian functions (SGFs) which are the curved-space analogs of cartesian Gaussian functions.
At low densities, the electrons become relatively localized into Wigner crystals, and the natural SGF centers are found by solving the Thomson problem (i.e. the minimum-energy arrangement of $n$ point charges) on the 3-sphere for various values of $n$.  We have found 11 special values of $n$ whose Thomson sites are equivalent.  Three of these are the vertices of four-dimensional Platonic solids --- the hyper-tetrahedron ($n=5$), the hyper-octahedron ($n=8$) and the 24-cell ($n=24$) --- and a fourth is a highly symmetric structure ($n=13$) which has not previously been reported.  By calculating the harmonic frequencies of the electrons around their equilibrium positions, we also find the first-order vibrational corrections to the Thomson energy.
\end{abstract}
%==========

\maketitle

%=================
\section{\label{sec:intro}
Introduction}
%=================
In a recent series of papers, \cite{EcLimit09, QuasiExact09, EcProof10, Glomium11, UEGs12, Ringium13, gLDA14, Wirium14} we have shown that the behavior of electrons in the (flat) Euclidean space $\mathbb{R}^D$ is surprisingly similar to the behavior in the (curved) $D$-dimensional manifold $\mathbb{S}^D$, the surface of a $(D+1)$-dimensional ball. \footnote{A $D$-sphere of radius $R$ is defined as the set of points in a $(D + 1)$-dimensional Euclidean space which are at distance $R$ from the origin.}  By exploiting this similarity between electrons on a line \cite{1DChem, Chem1D} and electrons on a ring \cite{QR12, Ringium13, gLDA14, Wirium14}, we have constructed a new type of correlation functional (the generalized local-density approximation) for density functional theory (DFT) calculations and we have shown that this new functional yields accurate correlation energies in a variety of one-dimensional systems. \cite{gLDA14, Wirium14}

However, our ultimate goal is to construct improved functionals \cite{Omega06} for three-dimensional (3D) systems and, to this end, we seek accurate correlation energies from the spherical space $\mathbb{S}^3$ (henceforth called a glome), which is the surface of a four-dimensional (4D) ball.  Electrons in $\mathbb{R}^3$ or $\mathbb{S}^3$ enjoy three degrees of translational freedom but the properties of $n$ electrons on a glome ($n$-glomium) have hitherto received little attention. \cite{TEOAS09, Concentric10, Glomium11, Exciton12, SGF14} 

An $n$-glomium is defined by the number $n$ of electrons and the glome radius $R$.  Its electron density is
\begin{equation}
	\rho = n / (2\pi^2 R^3),
\end{equation}
but it is often measured via the Wigner-Seitz radius \cite{VignaleBook}
\begin{equation}
	r_s = \left(\frac{3\pi}{2n}\right)^{1/3}R,
\end{equation}
which measures the average distance between neighbouring electrons.
High-density systems (which are weakly correlated) have small $r_s$ values while low-density systems (which are strongly correlated) have large $r_s$ values.  In the present study, we focus our attention on low-density glomiums with $2\le n \le 48$. 

In the low-density regime, the Coulomb energy (which decays as $R^{-2}$) dominates over the kinetic energy (which decays as $R^{-1}$) and the $n$ electrons localize onto \alert{particular points} on the glome that minimize their (classical) Coulomb repulsion.  \alert{These minimum-energy configurations} are called Wigner crystals \cite{Wigner34} and, if all of its sites are topologically equivalent, we will call it a uniform lattice. 

The work is organized as follows.  In Sec.~\ref{sec:E0}, we study the Thomson problem on a glome and discuss the uniform solutions.  In Sec.~\ref{sec:E1}, we calculate the harmonic vibrational energy of the electrons as they oscillate around their lattice positions.  In Secs.~\ref{sec:HF} and \ref{sec:VMC}, respectively, we report Hartree-Fock, near-exact and correlation energies of $n$-glomium at various densities.  Unless otherwise stated, all energies are reduced (i.e. per electron).
Atomic units are used throughout.

%%% TABLE 1 %%%
\begin{table*}
	\caption{\label{tab:E0} Non-reduced Thomson energy $\varepsilon_0$ and principal moments of inertia $I_k$ of the Thomson lattices on a unit glome.}
	\begin{ruledtabular}
	\begin{tabular}{clcccc}
	$n$			&	$\varepsilon_0$ 																									&	$I_1$	&	$I_2$	&	$I_3$	&	$I_4$	\\
	\hline
	2			&	$\ds{1/2}$																									&	$0$			&	$2$			&	$2$			&	$2$			\\
	3			&	$\ds{3/\sqrt{3}}$																							&	$3/2$		&	$3/2$		&	$3$			&	$3$			\\
	4			&	$\ds{6/\sqrt{8/3}}$																							&	$8/3$		&	$8/3$		&	$8/3$		&	$4$			\\
	5			&	$\ds{10/\sqrt{5/2}}$																							&	$15/4$		&	$15/4$		&	$15/4$		&	$15/4$		\\
	6			&	$\ds{9/\sqrt{2} + 6/\sqrt{3}}$																					&	$9/2$		&	$9/2$		&	$9/2$		&	$9/2$		\\
	8			&	$\ds{24/\sqrt{2} + 4/2}$																						&	$6$			&	$6$			&	$6$			&	$6$			\\
	10			&	$\ds{\frac{10}{2\sin(\pi/5)} + \frac{25}{\sqrt{2}}+ \frac{10}{2\sin(2\pi/5)}}$													&	$15/2$		&	$15/2$		&	$15/2$		&	$15/2$		\\[3mm]
	12\footnotemark[1]			&	$\ds{\frac{6}{2c} + \frac{12}{\sqrt{3}s} + \frac{12}{\sqrt{2+2s^2}}+ \frac{24}{\sqrt{2-s^2}} + \frac{12}{\sqrt{4-s^2}}}$					&	$6(1+c^2)$	&	$6(1+c^2)$	&	$6(1+s^2)$	&	$6(1+s^2)$	\\[3mm]
	13			&	$\ds{\sum_{k=1,2,4} \frac{26}{\sqrt{2-2\cos(k\pi/13)\cos(5k\pi/13)}}}$														&	$39/4$		&	$39/4$		&	$39/4$		&	$39/4$		\\[3mm]
	24			&	$\ds{96/1 + 72/\sqrt{2}+ 96/\sqrt{3} + 12/2}$																		&	$18$			&	$18$			&	$18$			&	$18$			\\[3mm]
	\mr{1}{*}{48}	&	$\ds{\frac{24}{2} + \frac{240}{\sqrt{2}} + \frac{48}{\sqrt{2\pm\sqrt{2}}}+ \frac{96}{\sqrt{2\pm\sqrt{2}/2}}+ \frac{96}{\sqrt{2\pm\sqrt{6}/2}}}$	&	\mr{1}{*}{$36$}	&	\mr{1}{*}{$36$}	&	\mr{1}{*}{$36$}	&	\mr{1}{*}{$36$}	\\[3mm]
				&	$\qquad\qquad \qquad \ds{+ \frac{96}{\sqrt{2\pm(\sqrt{3}-1)/2}}+ \frac{96}{\sqrt{2\pm(\sqrt{3}+1)/2}}}$																												\\[3mm]
	\end{tabular}
	\end{ruledtabular}
	\footnotetext[1]{$c = \cos\theta$, $s = \sin\theta$ and $\theta = 0.7935536685\ldots$}
\end{table*}

\begingroup
\squeezetable
\begin{table}
	\caption{
		\label{tab:energies}
		Reduced Thomson energy $E_0$, harmonic vibrational energy $E_1$, near-exact energy $E$, Hartree-Fock energy $\EHF$ and correlation energy $\Ec$ (in \mEh) for various $n$ and $r_s$
		}
	\begin{ruledtabular}
	\begin{tabular}{cccccc}
		\mc{2}{c}{$r_s$}		&	20			&	50			&	100			&	150			\\
		\hline
				& $E_0$		&	16.633		&	6.653		&	3.327		&	2.218		\\
				& $E_0+E_1$	&	23.068		&	8.281		&	3.902		&	2.531		\\
		$n=2$	& $E$		&	23.928		&	8.385		&	3.924		&	2.540		\\
				& $\EHF$		&	24.911		&	8.795		&	4.100		&	2.643		\\
				& $-\Ec$		&	0.983		&	0.410		&	0.176		&	0.103		\\
		\hline
				& $E_0$		&	33.557		&	13.423		&	6.711		&	4.474		\\
				& $E_0+E_1$	&	41.074		&	15.324		&	7.384		&	4.840		\\
		$n=3$	& $E$		&	41.783		&	15.405		&	7.400		&	4.847		\\
				& $\EHF$		&	43.811		&	16.079		&	7.664		&	5.000		\\
				& $-\Ec$		&	  2.028		&	  0.674		&	0.264		&	0.153		\\
		\hline
				& $E_0$		&	48.507		&	19.403		&	  9.701		&	6.468		\\
				& $E_0+E_1$	&	57.190		&	21.600		&	10.478		&	6.890		\\
		$n=4$	& $E$		&	57.155		&	21.550		&	10.461		&	6.882		\\
				& $\EHF$		&	59.886		&	22.358		&	10.757		&	7.046		\\
				& $-\Ec$		&	  2.731		&	  0.808		&	  0.296		&	0.164		\\
		\hline
				& $E_0$		&	62.009		&	24.804		&	12.402		&	8.268		\\
				& $E_0+E_1$	&	71.916		&	27.310		&	13.288		&	8.750		\\
		$n=5$	& $E$		&	71.038		&	27.119		&	13.238		&	8.728		\\
				& $\EHF$		&	74.240		&	27.985		&	13.537		&	8.888		\\
				& $-\Ec$		&	 3.202		&	  0.866		&	  0.299		&	0.160		\\
		\hline
				& $E_0$		&	75.564		&	30.226		&	15.113		&	10.075		\\
				& $E_0+E_1$	&	85.823		&	32.821		&	16.030		&	10.575		\\
		$n=6$	& $E$		&	85.406		&	32.682		&	15.986		&	10.553		\\
				& $\EHF$		&	88.283		&	33.530		&	16.293		&	10.720		\\
				& $-\Ec$		&	  2.877		&	  0.848		&	  0.307		&	  0.167		\\
		\hline
				& $E_0$		&	99.390		&	39.756		&	19.878		&	13.252		\\
				& $E_0+E_1$	&	110.951		&	42.681		&	20.912		&	13.815		\\
		$n=8$	& $E$		&	110.614		&	42.620		&	20.915		&	13.823		\\
				& $\EHF$		&	112.893		&	43.234		&	21.115		&	13.927		\\
				& $-\Ec$		&	    2.279		&	  0.614		&	  0.200		&	  0.104		\\
		\hline
				& $E_0$		&	122.336		&	48.934		&	24.467		&	16.311		\\
				& $E_0+E_1$	&	133.974		&	51.879		&	25.508		&	16.878		\\
		$n=10$	& $E$		&	133.522		&	51.767		&	25.485		&	16.874		\\
				& $\EHF$		&	136.154		&	52.479		&	25.727 		&	17.002  		\\
				& $-\Ec$		&	    2.632		&	  0.712		&	  0.242		&	  0.128		\\
		\hline
				& $E_0$		&	143.339		&	57.336		&	28.668		&	19.112		\\
				& $E_0+E_1$	&	155.615		&	60.441		&	29.766		&	19.710		\\
		$n=12$	& $E$		&	154.713		&	60.269		&	29.740		&	19.707		\\
				& $\EHF$		&	157.598		&	61.001		&	29.970		&	19.822		\\
				& $-\Ec$		&	    2.885		&	  0.732		&	  0.230		&	  0.115		\\
		\hline
				& $E_0$		&	153.600		&	61.440		&	30.720		&	20.480		\\
				& $E_0+E_1$	&	165.909		&	64.554		&	31.821		&	21.079		\\
		$n=13$	& $E$		&	164.804		&	64.322		&	31.767		&	21.061		\\
				& $\EHF$		&	167.947		&	65.126		&	32.033		&	21.196		\\
				& $-\Ec$		&	    3.143		&	  0.804		&	  0.266		&	  0.135		\\
		\hline
				& $E_0$		&	252.272		&	100.909		&	50.454		&	33.636		\\
				& $E_0+E_1$	&	265.600		&	104.280		&	51.647		&	34.285		\\
		$n=24$	& $E$		&	264.917		&	104.138		&	51.624		&	34.280		\\
				& $\EHF$		&	267.464		&	104.793		&	51.831		&	34.387		\\
				& $-\Ec$		&	    2.547		&	    0.655		&	  0.207		&	  0.107		\\
		\hline
				& $E_0$		&	425.792		&	170.317		&	85.158		&	56.772		\\
				& $E_0+E_1$	&	439.690		&	173.833		&	86.401		&	57.449		\\						
		$n=48$	& $E$		&	438.667		&	173.681		&	86.406		&	57.474		\\
				& $\EHF$		&	441.542		&	174.334		&	86.581		&	57.547		\\
				& $-\Ec$		&	    2.875		&	    0.653  		&	  0.175		&	  0.073		\\		
		\hline
%		$n=\infty$	& $-E$		&	31.316		&	14.450		&	7.677		&	5.269		\\
%				& $\EHF$		&	24.477		&	10.843		&	5.597		&	3.770		\\
		\alert{$n=\infty$}	& $-\Ec$		&	6.839   		&	 3.607     		&	2.080  		&	1.499		\\	
		\end{tabular}
	\end{ruledtabular}
\end{table}
\endgroup

%=================
\section{\label{sec:E0}
The Thomson problem}
%=================
What arrangement of $n$ unit point charges on a unit $D$-sphere minimizes their classical Coulomb energy?  This generalizes a question posed by J. J. Thomson as he devised the ``plum pudding'' model of atomic structure. \cite{Thomson04} Although the model was abandoned long ago, the Thomson problem continues to intrigue mathematicians and has resurfaced in many fields of science: surface ordering of liquid metal drops confined in Paul traps, \cite{Davis97} fullerenes-like molecules, \cite{Kroto85} arrangements of protein submits on spherical viruses \cite{Caspar62} or multielectron bubbles in liquid helium. \cite{Leiderer93, Tempere07}

The Thomson problem on a 1-sphere (i.e.~a ring) is trivial and the solutions consist of charges uniformly spaced around the ring.  The problem on a 2-sphere is challenging and, although it has been studied numerically up to large values of $n$, \cite{Erber97, Wales06, Wales09} mathematically rigorous solutions \cite{Schwartz13} have been established only for $n \in \{2, 3, 4, 5, 6, 12\}$.

The Thomson problem on a 3-sphere (i.e.~a glome) seeks the global minimum $\varepsilon_{0} = V(\bx_0)$ of
\begin{equation}
	V(\bx) = \sum_{i<j}^n r_{ij}^{-1},
\end{equation}
where $\bx$ describes the positions of the $n$ charges on the glome and $r_{ij}$ is the Euclidean distance between charges $i$ and $j$, measured through the unit glome.  It has attracted much less attention \cite{Altschuler07, Roth07} than the $D=2$ problem.

We performed a numerical study to determine the values of $n \le 50$ for which the Thomson \alert{minimum-energy configuration} on a glome is uniform.  \alert{Although it would be more suitable to use a global optimization, because we consider relatively small numbers of electrons, we adopt the following computational strategy:} for each $n$, we generated randomly at least 1000 distinct initial structures and minimized their energy using local optimization algorithms, as implemented in the \textsc{Mathematica} software package. \cite{Mathematica10}  Our numerical experiments indicate that there are eleven uniform lattices.  Their energy $\varepsilon_0$ and principal moments of inertia are listed in Table \ref{tab:E0} and their cartesian coordinates are in supplementary material. \footnote{See supplemental material at [URL will be inserted by AIP] for the cartesian coordinates of the Thomson lattices.}  The set $\{2, 3, 4, 6, 8, 12, 24\}$ of $n$ values for uniform $D=2$ Thomson lattices \cite{SGF14} is a subset of the $D=3$ set but we do not understand this.

The principal moments of inertia of a lattice indicate the degree of its symmetry.  Generalizing the standard notation for 3D structures, \cite{SimonsBook} we define  a ``hyperspherical top'' as a 4D structure with four equal moments;  the extremely symmetrical $n = 5$, 6, 8, 10, 13, 24 and 48 lattices are of this type.
We define a ``spherical top'' as a lattice in which three of the four moments are equal;  the $n=2$ (prolate) and $n =4$ (oblate) lattices are of this type.
We define a ``symmetric top''  as a lattice in which the moments form two pairs; the $n = 3$ and $n=12$ lattices are of this type.

The glome lattices for $n = 2$ (a diameter) and $n = 3$ (equilateral triangle) are the same as on a 1-sphere and 2-sphere.  The glome lattice for $n = 4$ (regular tetrahedron) is the same as on a 2-sphere. 
The $n = 5$ lattice is a regular hyper-tetrahedron (also called a regular simplex\cite{CoxeterBook}), a 4D Platonic solid with ten equal side lengths.
The $n = 6$ lattice is the union of an equilateral triangle in the $wx$-plane and another such triangle in the $yz$-plane.
The $n = 8$ lattice is a hyper-octahedron (or 16-cell), a 4D Platonic solid with vertices at $\pm1$ on each of the four cartesian axes.
The $n = 10$ lattice is the union of a regular pentagon in the $wx$-plane and another such pentagon in the $yz$-plane, while the $n = 12$ lattice is the union of two perpendicular triangular prisms.
The $n = 13$ lattice is peculiar to the $D=3$ Thomson problem and, to the best of our knowledge, has not been previously described. 
The $n = 24$ lattice is the 24-cell, a 4D Platonic solid with no analogue in 3D.  
It is the union of a hyper-octahedron and a hyper-cube.
The $n = 48$ lattice is peculiar to the $D=3$ Thomson problem.

%=====================
\section{\label{sec:E1}
Harmonic vibrational energy}
%=====================
The energy $E_\text{W}$ of an $n$-electron Wigner crystal on a glome can be estimated by solving the Schr\"odinger equation in the harmonic potential
\begin{equation}
	V_2(\bx) = V(\bx_0) + \frac{1}{2} (\bx-\bx_0)^\text{T} \cdot \mathbf{H} \cdot (\bx-\bx_0),
\end{equation}
where $\mathbf{H}$ is the $3n \times 3n$ second-derivative (Hessian) matrix 
\begin{equation}
	H_{ij} = \left. \frac{\partial^2 V(\bx)}{\partial t_i \partial t_j}\right|_{\bx=\bx_0},
\end{equation}
and the $t_i$ are suitable tangential coordinates.  The square roots of the Hessian eigenvalues are the harmonic frequencies $\omega_i$ and one can then write
\begin{align}
\label{eq:EW}
	E_\text{W}	& = \frac{\varepsilon_0}{R} + \frac{\sum_{i=1}^{N_\text{vib}}\omega_i}{2R^{3/2}} + O(R^{-2})	\\
				& = E_0 + E_1 + O(R^{-2}).
\end{align}
One finds that, for $n > 3$ particles in $\mathbb{S}^3$, exactly $3n - 6$ of the Hessian eigenvalues are non-zero and six vanish because they correspond to rotations on the glome.  This is analogous to the familiar $3n - 6$ rule \cite{WilsonBook} for non-linear molecules vibrating in $\mathbb{R}^3$.  Numerical values of $E_0 + E_1$ for a range of $n$ and $r_s$ are presented in Table \ref{tab:energies}.

%================
\section{\label{sec:HF}
Hartree-Fock energies}
%================
We now turn to the ab initio treatment of $n$ spin-up electrons on a glome, i.e.~ferromagnetic $n$-glomium.  We have performed Hartree-Fock (HF) calculations\cite{SzaboBook} in a basis of $s$-type spherical Gaussian functions (SGFs) \cite{SGF14}
\begin{equation}
	G_\alpha^{\bm{A}} (\bm{r}) = \sqrt{\frac{\alpha}{2\pi^2 I_1(2\alpha)}}\exp(\alpha\,\bm{r} \cdot \bm{A}),	\quad	 \bm{r} \in \mathbb{S}^3,
\end{equation}
where $\bm{A} \in \mathbb{S}^3$ is the center of the SGF, $\alpha$ is the exponent and $I_1$ is a modified Bessel function. \cite{NISTbook}  An SGF behaves like a Gaussian near $\bm{A}$ and is therefore a natural basis function for describing a localized electron.  Moreover, the product of two SGFs is a third SGF which make them computationally attractive. \cite{Review94}  All the required one- and two-electron integrals can be found in Ref.~\onlinecite{SGF14}.

\begin{figure}
	\includegraphics[width=0.45\textwidth]{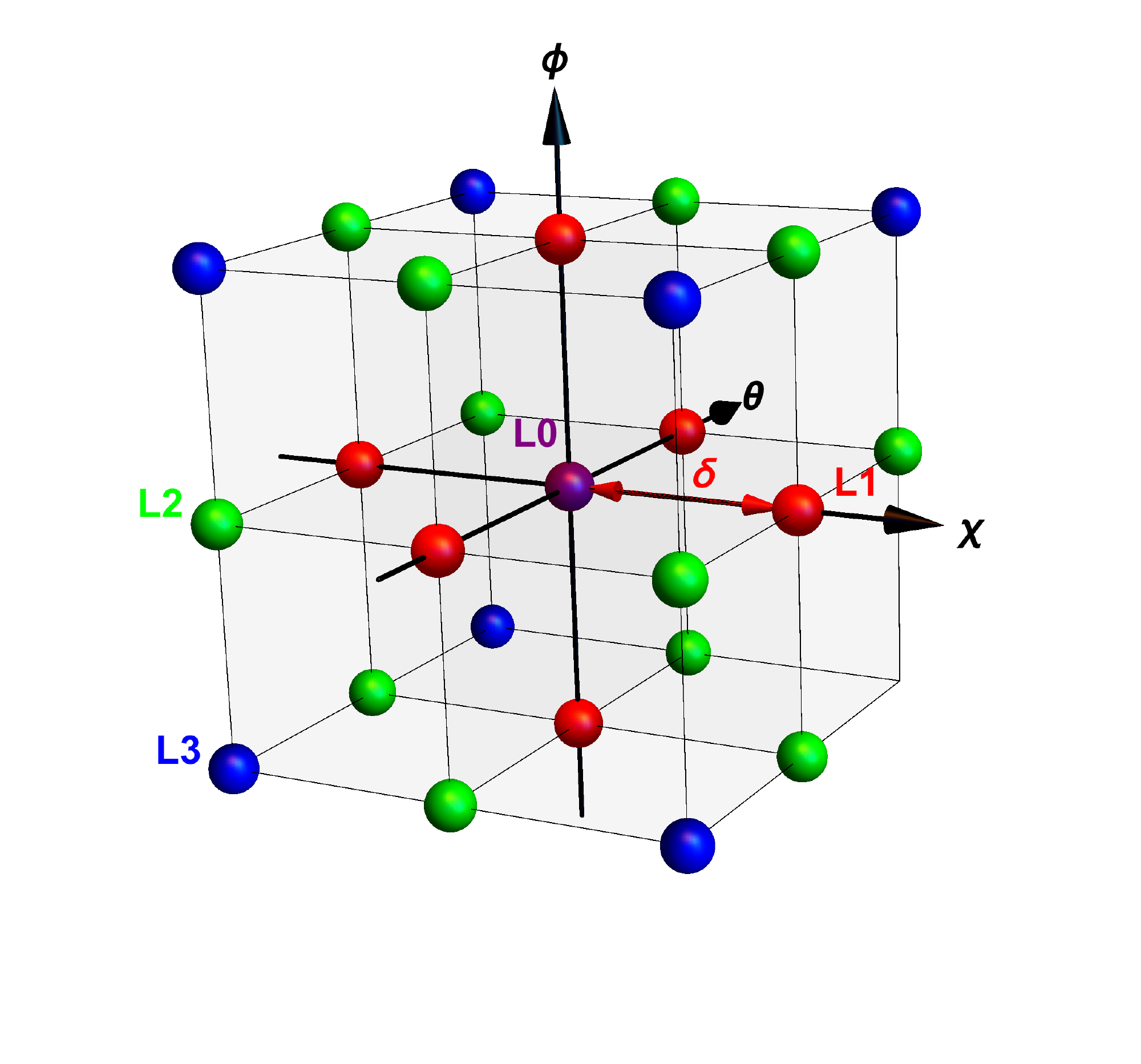}
	\caption{
	\label{fig:grid}
	Three-dimensional SGF grid used at each Thomson site in HF calculations. 
	The Level 0 (L0), Level 1 (L1), Level 2 (L2) and Level 3 (L3) functions are represented in purple, red, green and blue, respectively.
	$\delta$ is the shortest distance (around the glome) between two grid points.}
\end{figure}

\begin{table*}
	\caption{\label{tab:HFrs20}
		HF energies (in \mEh) for ferromagnetic $n$-glomium with $r_s = 20$.
		There are $M$ basis functions per Thomson site.}
	\begin{ruledtabular}
	\begin{tabular}{ccccccccccccc}
		\mc{2}{c}{Basis set}				&	\mc{11}{c}{Number of electrons $n$}	\\
		\cline{1-2}					\cline{3-13}																										
		Level	&	$M$		&	2		&	3		&	4		&	5		&	6		&	8		&	10		&	12		&	13		&	24		&	48		\\
		\hline
		0		&	1		&	24.983	&	43.939	&	60.016	&	74.277	&	88.345	&  112.911	&  136.249	&  157.626	&  167.991	& 267.468		& 441.543		\\
		1		&	7		&	24.917	&	43.816	&	59.890	&	74.253	&	88.290	&  112.899	&  136.169	&  157.606	&  167.951	& 267.464		& 441.542	 	\\
		2		&	19		&	24.911	&	43.811	&	59.886	&	74.244	&	88.286	&  ''			&  136.156	&  157.605	&  167.947	& ''			&	''	 	\\
		3		&	27		&	''		&	''		&	''		&	''		&	''		&  ''			&  ''			&  ''			&  ''			& ''			&	''	 	\\
	\end{tabular}
	\end{ruledtabular}
\end{table*}

Adopting the ``Gaussian lobe'' philosophy introduced by Whitten \cite{Whitten66} many years ago, we use off-center $s$-type SGFs to mimic SGFs of higher angular momentum.  The basis set consists of a grid of $s$-type SGFs with same exponent $\alpha$ clustered around each Thomson site (see Fig.~\ref{fig:grid}).  The complete basis set (CBS) HF energy is obtained by successively adding Level 0 (L0), Level 1 (L1), Level 2 (L2) and Level 3 (L3) functions to the basis set.  In each calculation, we optimize the SGF exponent $\alpha$ and the L0/L1 distance $\delta$ using the Newton-Raphson optimization procedure.  Our target accuracy was 1 microhartree (\uEh) per electron.  The resulting HF energies for a range of $n$ and $r_s$ are shown in Table \ref{tab:energies}.

Table \ref{tab:HFrs20} reports HF energies of $n$-glomium as the basis set is gradually improved.  SGFs are optimal for localized electronic systems but become less efficient as the density increases so the convergence for $r_s > 20$ is always at least as fast as for $r_s = 20$.  We therefore show results for $r_s = 20$, the most challenging case.

For a given value of $r_s$, the minimal-basis (L0) exponent $\alpha$ grows, i.e.~the electrons become more localized, as $n$ increases.  The results of Table \ref{tab:HFrs20} show that L2 achieves \uEh\ accuracy for all $n$ values and, indeed, L1 suffices for the largest $n$ values.  It is well known that, on a 2-sphere, the number of nearest neighbors around an electron approaches six (hexagonal lattice) for large $n$. \cite{Bowick02}  Similarly, on a glome, the number of nearest neighbors approaches eight (body-centered cubic lattice).\cite{VignaleBook}  Thus, for large $n$, the density around each electron becomes approximately isotropic, and the L2 and L3 functions become largely superfluous.

%=================
\section{\label{sec:VMC}
Near-exact energies}
%=================

To obtain near-exact energies for electrons on a glome, we have performed variational Monte Carlo (VMC) calculations. \cite{Umrigar99}  The trial wave function is of the form 
\begin{equation}
	\Psi_\text{T} = \Psi_0 \,e^J,
\end{equation}
where $\Psi_0$ is a Slater determinant of either SGFs \cite{SGF14} or hyperspherical harmonics \cite{Avery85, AveryBook} (HSHs)
\begin{equation}
	Y_{k \ell m}(\chi,\theta,\phi) = C_{k-\ell}^{\ell+1}(\cos\chi) \sin^\ell \chi \,Y_{\ell m}(\theta,\phi).
\end{equation}
$C_{k}^{\ell}$ is a Gegenbauer polynomial and $Y_{\ell m}$ is a spherical harmonic. \cite{NISTbook}  The Jastrow factor $J$ is a symmetric function of the interelectronic distances $r_{ij}$ containing two-body (2B), three-body (3B) and four-body (4B) terms. \cite{Nodes15, Huang97}  More details will be reported elsewhere. \cite{LoosQMC}

At low densities, the energy minimization procedure is unstable and the parameters of the Jastrow factor were therefore optimized by variance minimization using Newton's method. \cite{Umrigar05, Toulouse07, Umrigar07, Toulouse08}  In all calculations, the statistical error obtained by reblocking analysis \cite{CASINO10, Lee11} is under 1 \uEh. For small numbers of electrons, comparisons with extrapolated full configuration interaction (FCI) calculations \cite{Knowles84, Knowles89} indicate that our VMC energies have sub-\uEh\ accuracy.  They are reported in Table \ref{tab:energies} for various $r_s$ and $n$ values.

Because we have observed that many-body effects are more important for small $r_s$, we have studied the convergence of the energy for $r_s = 20$ and various $n$ values as 2B, 3B and \alert{4B} terms are successively included.  We found that the inclusion of 3B terms systematically lowers the energies by up to 20 \uEh\ per electron but that the inclusion of 4B terms offers less than 1 \uEh\ per electron.  We therefore eschewed 4B terms in the calculations with $r_s > 20$. 

When SGFs were used, the determinant $\Psi_0$ corresponds to a HF Level 0 calculation (i.e.~a single SGF on each Thomson site) but the value of the SGF exponent was optimized to minimize the VMC energy.  We found that the SGF basis is superior to the HSH basis for $n=8$, 10, 12 and 24, while the two basis sets yield identical energies for the other cases.  8-, 10-, 12- and 24-glomium are ``open-shell'' systems, i.e.~the highest occupied shell of HSHs is partially occupied and there are several low-lying determinants with significant weights.  In contrast, SGFs at the Thomson lattice sites naturally describe the localized electrons and are particularly well suited to these systems.  For 48-glomium, computational limitations precluded full exponent optimization and the results in Table \ref{tab:energies} were therefore obtained with HSHs.  They are probably less accurate than the other energies.

%==============
\section{Discussion}
%==============
By taking the difference between the CBS-HF energies of Sec.~\ref{sec:HF} and the VMC energies of Sec.~\ref{sec:VMC}, we have obtained the near-exact correlation energies $\Ec$ of $n$-glomium for various $r_s$ and $n$ values.  Our results are reported in Table \ref{tab:energies}.

\begin{figure}
	\includegraphics[width=0.45\textwidth]{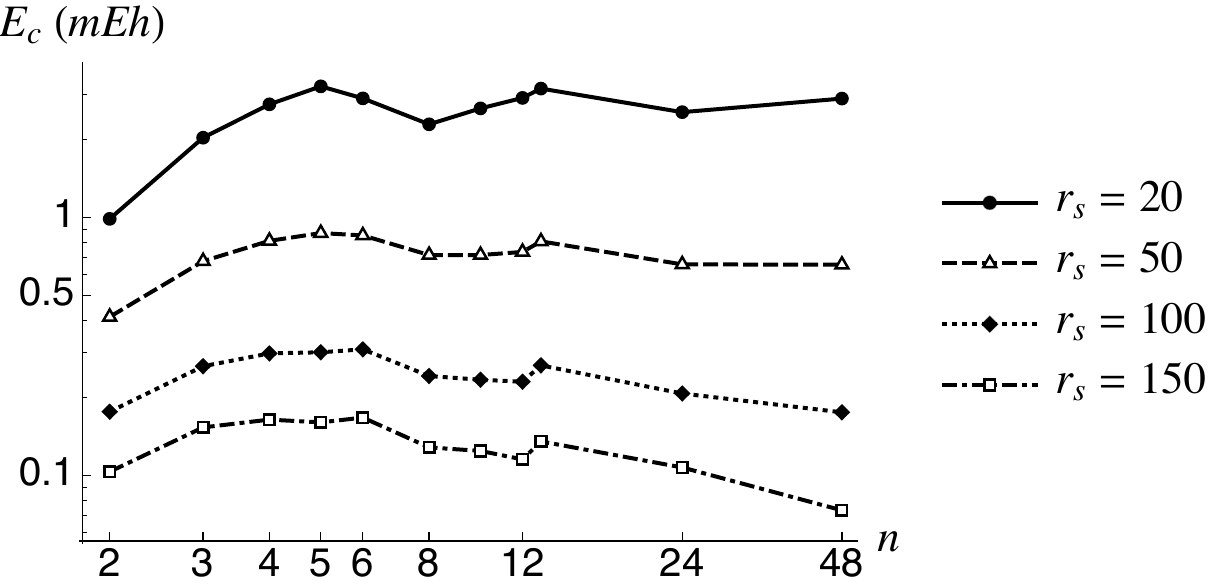}
	\caption{
	\label{fig:plotEc}
	Reduced correlation energy $\Ec$ (in \mEh) as a function of $n$ for various $r_s$.}
\end{figure}

While we have shown that, in 1D, the correlation energy is a smooth and monotonic function of $n$, \cite{Ringium13, gLDA14} the situation is rather different in 3D. 
As shown in Figure \ref{fig:plotEc} (where have plotted $\Ec$ as a function of $n$), the reduced correlation energies do not change monotonically as $n$ increases.  Instead, they initially increase and reach a maximum at $n = 5$ or $n = 6$.
Beyond $n = 5$, they oscillate and tend to decrease slowly with $n$, especially at very low densities. 
The oscillations are probably due to ``shell effects'' which originate from partially-filled energy level in open-shell systems (see above).
Such shell effects are also observed in 2D. \cite{SGF14}
For fixed $n$, our numerical results show that $\Ec$ decreases as $r_s^{-3/2}$ for large $r_s$.
This is expected due to the cancellation of the leading term (proportional to $r_s^{-1}$) in the exact and HF energies expansion at large $r_s$ (see Eq.~\eqref{eq:EW}).
\alert{We have also reported the correlation energies \cite{Drummond04, Ceperley02, Spink13} of the jellium model (which corresponds to $n\to\infty$) in Table \ref{tab:energies}.
Exact jellium energies are available in many previous papers \cite{Ceperley80, Kwon98, Ortiz99, Ceperley02, Trail03, Drummond04, Zhang08, Spink13}  but fully relaxed HF energies \cite{Overhauser59, Overhauser62} are rare.
As a result, most of the jellium correlation energies in the literature are based on unrelaxed HF energies and are consequently significantly larger than ours.}

The harmonically corrected Thomson energy ($E_0+E_1$) is usually higher than the exact energy but $n=2$ and $n=3$ are exceptional cases.  At very low densities, however, it always approximates the exact energy well.  Including the first anharmonic correction $E_2$ would probably yield even better estimates. \cite{Carr61a, Carr61b}

%--------------------------------------------------------------------
\section{Conclusion}
%--------------------------------------------------------------------
The goal of this work was to generate benchmark correlation energies for the development of improved correlation functionals for DFT calculations on 3D systems.  To achieve this, we have studied the correlation energies of low-density spin-polarized electron gases on a glome. 

First, we looked at the Thomson problem on a glome, using numerical optimization algorithms to locate uniform lattices, i.e. those in which all the lattice sites are equivalent, for $n \le 50$.  We found eleven uniform lattices.
We also noted that the structure of three of these uniform lattices correspond to well-known 4D Platonic solids: the hyper-tetrahedron ($n=5$), the hyper-octahedron ($n=8$) and the 24-cell ($n=24$). 
Moreover, we have pointed out the highly symmetric case of $n=13$ and we stressed that this polychoron has not been previously described anywhere in the literature. 
By taking into account the quantum oscillation of the electrons around their equilibrium positions, we obtained the harmonic vibrational contribution to the (classical) Thomson energy. 
As expected, the sum of the Thomson and harmonic energies is a very good approximation of the exact energy at very low density.

Moreover, by systematically increasing the number of $s$-type SGF basis functions around each Thomson site, we obtained the CBS HF energy of $n$-glomium for a range of densities ($20\le r_s\le 150$). 
In general, the convergence analysis reveals that only L2 calculations are required to converge the HF energies to microhartree accuracy. 
However, we note that as we move into the high-density regime, more $s$-type SGF functions will be needed to reach the CBS limit.
In this regime, HSHs might constitute a more suitable one-electron basis set. 
We will investigate this in a forthcoming paper.

The near-exact energies were obtained using highly-accurate VMC stochastic calculations using HSH and SGF one-electron basis sets.  We have shown that 4B terms in the Jastrow factor have insignificant effect on the energy.  The energies obtained using both basis sets are in very good agreement, except for 8-, 10-, 12- and 24-glomium where the SGF basis is superior.

The present work is a significant step towards the construction of correlation functionals for molecules and solids.  The next step is to generate accurate correlation energies within the high density regime, and for partially-polarized systems.  This work is underway and our results will be reported elsewhere.

%%%%%%%%%%%%%
\begin{acknowledgments}
P.F.L.~thanks Julien Toulouse for useful discussions.
P.M.W.G.~and P.F.L.~thank the NCI National Facility for generous grants of supercomputer time. 
P.M.W.G.~thanks the Australian Research Council for funding (Grants No.~DP120104740 and DP140104071). 
P.F.L. thanks the Australian Research Council for a Discovery Early Career Researcher Award (Grant No.~DE130101441) and a Discovery Project grant (DP140104071).
\end{acknowledgments}
%%%%%%%%%%%%%


\end{document}